\RequirePackage{fix-cm}
 \documentclass[a4paper, 12pt]{article}   
\usepackage{graphicx}
\usepackage{amssymb}

\begin{document}

\title{ Universe made of baryonic gravitating particles behaves as 
a $\Lambda$CDM Universe}


\author{Miguel Portilla\\
{\footnotesize Departament d'Astronom\'{\i}a i Astrof\'{\i}sica}\\
{\footnotesize Universitat de Val\`encia , Spain}\\
{\footnotesize {miguel.portilla@uv.es}}}



\maketitle

\begin{abstract}
Using an approximate  solution to the $N$-body problem in general relativity, and the \emph{principle of local isotropy at any point},  we construct a cosmological model, with zero curvature, for a universe composed uniquely by collision-less gravitating point-particles. The result is not, as currently thought, a null pressure Friedman model, but one that reproduces quite well the dark phenomena.

We assume that there exist three consecutive ages  with this property, formed  by free atoms,   stars and   galaxies, respectively. Certainly, we  are using  a highly idealized  view of the very  complicated   process going from uncoupled atoms to galaxies, but it allows us to obtain that 
the energy density at each epoch is of the form $  \rho(a)=\frac{3H_0^2}{8\pi G}\left(\frac{\Omega_{ba}(1+\alpha)}{a^3}+ f(a)\right) $ , where $\alpha$ is a constant, that we identify with the dark matter,  and
 $f(a)$  a function of the scale factor, which  is zero at galaxy formation and  \emph{practically constant at the present epoch}, constant that  we identify with the cosmological constant. 
 
 The parameters of our model are the baryonic density $ \Omega_{ba}$ and the redshifts $ z''_i , z'_i , z_i $, corresponding to the effective decoupling of atoms and radiation, the formation of stars and galaxies respectively. The model sets a relation between the galaxy formation epoch and the amount of dark matter and dark energy, e.g., galaxy formation at $z_i \approx 11$  produces $\Omega_{\Lambda}=0.683$; and the  function $f(a)$ predicts the begining  of the acceleration recently at redshift
  $z\approx 0.6 $, just as the $\Lambda CDM$ model.  So, the dark phenomena could be related to a revision of the dynamical description of a gas of collision-less gravitating particles.
\end{abstract}
\section{Introduction}

We want to show that dust is not exactly the continuous model for a universe formed by collision-less  gravitating particles, and that may be the clue for explaining both dark matter and dark energy. 
Dust is a model of matter described by a perfect fluid energy tensor without pressure nor internal energy density:  $T^{\mu \nu}=\mu u^{\mu}u^{\nu}$.  In cosmology this assumption produces the Einstein de Sitter cosmological model. Long-standing problems and more recently the acceleration manifested by high redshift supernovae have reinstated the cosmological constant to obtain a realistic cosmological model ($\Lambda$CDM). The metric of any Robertson Walker model (RW) is determined up to the scale factor $R(s)$, a function of the cosmological time $s$, and the curvature index $k=0,-1,1$;  and it is completely determined by giving the universe's  energy density as  function of the dimensionless expansion factor  $a(s)=R(s)/R_o $, with $R_o$ the present value of the scale factor. We shall need an expression of the form
$  \rho(a)=\frac{3H_0^2}{8\pi G}\left(\frac{\Omega_m}{a^3}+f(a)\right) $ 
, where  $\Omega_m = \Omega_{ba}+ \Omega_{dm}$ is the rest mass contribution composed by baryonic and dark matter and $f(a)$ any other contribution. The scale factor is obtained by solving the Friedman equation 
$\frac{da}{ds}=H_{0}\sqrt{\frac{\Omega_{m}}{a}+ \Omega_{k}+a^{2}f(a)} $, where  $\Omega_{k}=-k/H_{0}^{2}R_{0}^{2}$ is the curvature of the equal time sections (divided by the critical density). 
To date, all the  observations make  the case for  the  so called concordance cosmological model, that assumes zero curvature and accepts the cosmological constant taking  $f(a)=  \Omega_{\Lambda}$. So the previous equation reduces to:
\begin{equation}
\frac{da}{ds}=H_{0}\sqrt{\frac{\Omega_{m}}{a}+a^{2}\Omega_{\Lambda}} 
\end{equation}
 Taking the last PLANCK's mission parameter estimations
 \begin{equation}
 \left(\Omega_{ba} , \Omega_{dm} , \Omega_{\Lambda}\right)=\left(0.049 , 0.268 ,  0.683\right)
 \end{equation}
  one  concludes that the universe needs  baryonic matter, and unknown dark matter and  dark energy.
We will show how we can use baryonic matter alone to construct a cosmological model approximating  the concordance model quite well. We need for that to examine the problem of collision-less particles in  the cosmological context.

\section{\label{S:Dyna}Dynamics of  gravitating particles}
Any space time $(\sf M,g,T)$  may be expanded as an infinite series $g=\eta+\sum_{m=1}^{\infty}g^{(m)}$, $T=\sum_{m=0}^{\infty}T^{(m)}$, with the Minkowski metric as approximation of  order zero, and $g^{(m)}=O(G^{m})$, and $T^{(m)}=O(G^m)$;  an appropriate dimensionless parameter will be introduced  later. The following procedure  is the only known way to determine the motion of a finite number of auto gravitating particles. We shall follow the variant  \cite{HG} that describes the energy tensor by distributions 
\begin{equation}
\sqrt{-g}T^{\rho \sigma}=\sum_{j=1}^{N}\int M_{j}(\tau_{j})v_{j}^{\rho}v_{j}^{\sigma} \delta^{(4)}(x^{\rho}-z_{j}^{\rho}(\tau_{j})) d\tau_{j}, 
\end{equation}\label{E:T} 
where $v_{j}^{\rho}=dz_{j}^{\rho}/d\tau_{j}  $ and  $d\tau_{j}=(\eta_{\alpha \beta}dz_{j}^{\alpha} dz_{j}^{\beta})^{1/2}$.
The Minkowski metric, with signature $(+,-,-,-)$, is used as an auxiliary metric for parametrize the world lines of the particles and for introducing  coordinates $(x^0=t,x^{a})$  fixed, up to arbitrary Lorentz transformations, by the coordinate conditions $  \eta^{\nu \lambda} \partial_{\lambda}\left(g_{\mu \nu}-\frac{1}{2}\eta_{\mu \nu}\eta^{\alpha \beta}g_{\alpha \beta}\right)=0$. 

By writing  the field equations in the form $\sqrt{-g}G^{\mu}_{\nu} = - 8\pi G \sqrt{-g}T^{\mu}_{\nu}$ and  introducing the  Lorentz covariant tensors
\begin{eqnarray}
\overline T_{\mu \nu}^{(m)}&=&(\sqrt{-g}T^{\alpha}_{\mu})^{(m)}\eta_{\alpha \nu} 	\  , \  \  \gamma^{(m)}_{\mu \nu}=g_{\mu \nu}^{(m)}-\frac{1}{2}\eta_{\mu \nu}\eta^{\alpha \beta}g^{(m)}_{\alpha \beta} \ ,\nonumber \\ && g^{(m)}_{\mu \nu}=\gamma_{\mu \nu}^{(m)}-\frac{1}{2}\eta_{\mu \nu}\eta^{\alpha \beta}\gamma^{(m)}_{\alpha \beta},
\end{eqnarray}
Havas and Goldberg (HG) \cite{HG} expanded  Einstein's equations and coordinate conditions in powers of $G$:

\begin{equation}\label{E:field}
\sum_{m=1}^{n}(\Box \gamma_{\mu \nu}^{(m)}+2\Lambda_{\mu \nu}^{(m)})=
-16 \pi G \sum_{m=0}^{n-1}\overline T_{\mu \nu}^{(m)}
\end{equation}

\begin{equation}\label{E:cc}
\sum_{m=1}^{n}  \eta^{\nu \lambda} \partial_{\lambda} \gamma^{(m)}_{\mu \nu} = 0,
\end{equation}
where  $\Lambda_{\mu \nu}^{(m)}$ contains only terms nonlinear in  the $\gamma^{(r)}_{\mu\nu}$ of orders $r\leq m-1$. Let us remark that we have changed the index notation  used by HG  in the expansion of the energy tensor. Our super index in $T^{(m)}$ runs from zero (instead of from one) to infinity. So, $T^{(0)}$ indicates that it is independent of $G$.

The $nth$ order approximation for the metric is obtained as a functional of particle motion, but the parametrization $z_{j}^{\rho}(\tau)$ should be obtained by solving the equations of motion up to order $n$, which are  derived, as well as the mass  $M_j^{(n)}$ to order $n$,  as integrability conditions on the field equations in the $(n+1)$ order approximation  \cite{HG}. The motion can also be obtained  by requiring the coordinate conditions up to order $n$. To order zero  one gets uniform motions  and constant masses. The mass $M_{j}$ is then obtained as a series $M_{j}=\sum_{m=0}^{\infty}M_{j}^{(m)}$, with $M_j^{(0)}=m_j=$constant .
Consequently one can determine both the energy tensor and the metric in a recurrent way: 
\begin{equation}\label{E:sequence}
 g^{(0)}=\eta \rightarrow \overline T^{(0)}\rightarrow g^{(1)}\rightarrow \overline T^{(1)}\rightarrow g^{(2)}... 
\end{equation}
Thus one obtains a sequence of space-times 
\begin{equation}
(\sf M, \eta ,0),(\sf M, \eta + g^{(1)} ,\overline T^{(0)})  ,  (\sf M, \eta + g^{(1)}+g^{(2)} ,\overline T^{(0)}+\overline T^{(1)}),...
\end{equation}
which one assumes converge to a real space-time.

\section{ Gravitating particles in the cosmological context}

To deal with this issue in the cosmological context we shall assume \emph{the principle of local isotropy at every point}. That assumtion makes the calculation of the metric up to the first order very simple, and it is well known  \cite{RDLER} that that hypothesis by itself is enough to derive a  RW   cosmological model. So, in section~\ref{S:RW} we shall associate a $RW$ model to each espace time of the foregoing sequence. But, first of all we must  study two basic problems. One is to translate the discrete nature of the particles, described by Dirac distributions,  into a fluid described by continuous functions. This will be done in section~\ref{S:MET}. The other problem (considered here in subsection ~\ref{SS:3ages})  is to identify the particles: we have assumed that, to date, we have had three eras dominated by free gravitating particles. The first was formed by atoms freed of  the electromagnetic perturbations, then the atoms collapsed into stars, and later the stars collapsed into galaxies. Hence, the model of universe we are going to describe will introduce  four parameters $(\Omega_{ba}, z''_i, z'_i, z_i)$:  baryonic density,  redshifts of definitive decoupling of photons and atoms , the birth of stars,  and the birth of galaxies respectively.

\subsection{Metric and mass up to the first order for the phase of galaxy dominance}

\subsubsection{The zero order}

\begin{figure}[hbt]\label{Milne}
\centering
\includegraphics{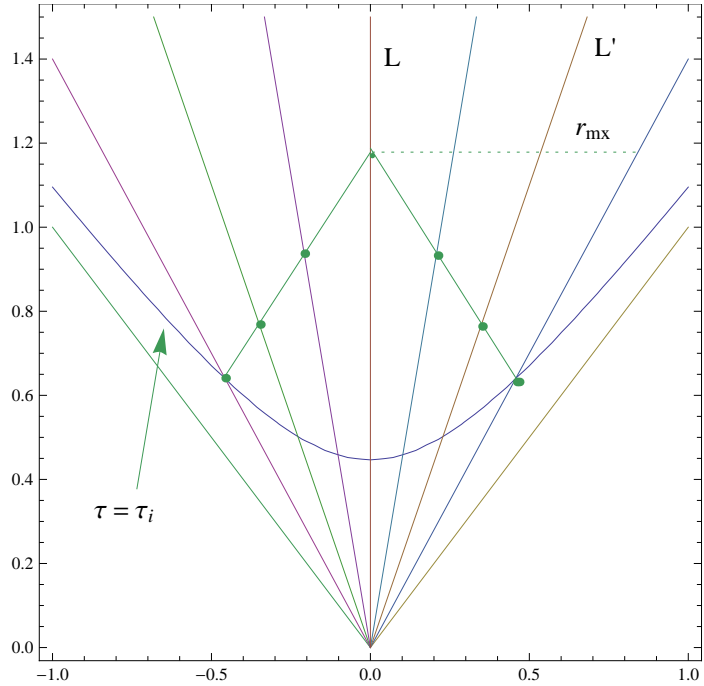}
\caption{Diagram of the Milne universe, showing the world lines of the galaxies and the past light cone of a point over the observer $L$. This figure is important to understand what follows. As the event shown on line $L$ moves upwards, the number of particles interacting with $L$ increases and the number density decreases. This is the clue for producing dark phenomena.\label{Fi:Milne} }
\end{figure}

At zero order in $G$, i.e.,  no energy at all,  we must consider the Milne universe \cite{RDLER} : we draw a future light cone in Minkowski space and infinite free particles world lines satisfying the isotropy hypothesis, as shown in  Figure~\ref{Fi:Milne}.  The world line passing through the point $(r,t)$ has 4-velocity $u^{(0)}=\gamma^{(0)}(\partial t + (r/t)  \partial r)$ with  relativistic factor $\gamma^{(0)}=t(t^2-r^2)^{-1/2}$ with respect to one observer of reference L. All observers are equivalent for they are connected by a Lorentz boost. They should be distributed with number density $n^{(0)}(r,t)=\gamma \rm N \tau^{-3}  = \rm N t (t^2-r^2)^{-2}$, with $\rm N$ a constant. The number density is constant on the hyperbolas  of  constant  proper time  $t^2-r^2 = const.$ , which are Minkowski orthogonal to the particle's world lines. The past light cone of an event on the reference world line $(0, t > \tau_i )$ intersects the hyperbola $\tau = \tau_i$, representing the birth of the galaxies, defining the radial coordinate  $r_{max}=t \frac{t^2-t_i^2}{t^2+t_i^2}$ shown in Figure~\ref{Fi:Milne}.  

\subsubsection{\label{SS:order1}The  first order}
The first order approximation of the metric is obtained assuming that the  particles (say galaxies) were born at some definite proper time $\tau_i$. This assumption allows us to  consider at any instant only a finite number of particles. Otherwise the problem, with infinite particles, would be nonsensical. As stated in section~\ref{S:Dyna},  integrability conditions  on the first order  field equations   determine at zero order  constant masses  $M^{(0)}_j= m_j$, and uniform motions for the particles, that is, they determine   completely the zero order energy tensor. \emph{We identify the constant masses  $m_j$ with the baryonic rest mass of the galaxies}. 

The  zero order energy tensor determines the metric up to the  first order in $G$. Taking null initial conditions on the surface $ \tau = \tau_i$ we have  \cite{HG} 
\begin{equation}\label{E:g1}
 g^{(1)}_{\mu \nu}=-4G m \sum_{j=1}^{n} \frac{(\eta_{\mu \alpha}\eta_{\nu \beta}v_j^{\alpha}v_j^{\beta}-\frac{1}{2}\eta_{\mu \nu})_r}{\eta(x-z_j,v_j)_r},
\end{equation}
where we consider all the particles with the same mass,  the subscript $r$ means evaluated at the retarded event, and the sum is over all the particles whose world line intercepts the observer's past light cone, as shown in Figure \ref{Fi:Milne}. It is singular at the particles world lines.

With the metric up to the first order, integrability conditions  on the  field equations up to second order  determine the first order mass correction  
\begin{equation}\label{E:M0}
M^{(1)}(t) = - \frac{1}{2}m g^{(1)}_{\alpha \beta}v^{\alpha}v^{\beta} + C^{(1)}, 
\end{equation}
where $C^{(1)}$ is a constant of integration, and the equations of motion up to the first order 
\begin{eqnarray}\label{E:em1}   
&&m \frac{d}{d\tau}\left(\left(\eta_{\mu\rho}+g^{(1)}_{\mu \rho}\right)v^{\rho}-\left(\frac{1}{2}g^{(1)}_{\alpha \beta}v^{\alpha}v^{\beta} - C^{(1)}\right)\eta_{\mu\rho}v^{\rho}\right)=\nonumber \\
&&\frac{1}{2}m \partial_{z^{\mu}}g^{(1)}_{\alpha \beta}v^{\alpha}v^{\beta}.
\end{eqnarray}

 Dealing with the first order in $G$, the divergences over the world lines may be treated as in electromagnetism by an appropriate choice of the constant $C^{(1)}$, obtaining for the metric over any particle, namely particle $k$, a finite expression:
\begin{eqnarray}\label{E:g2} 
&& g^{(1)}_{\mu \nu}(z_k(\tau))= -4G m \sum_{j\neq k}^{n} \frac{(\eta_{\mu \alpha}\eta_{\nu \beta}v_j^{\alpha}v_j^{\beta}-\frac{1}{2}\eta_{\mu \nu})_r}{\eta(x-z_j,v_j)_r}\nonumber\\
&& +4m(\dot v_k^{\alpha} v_k^{\beta}+v_k^{\alpha} \dot v_k^{\beta})\eta_{\mu\alpha}\eta_{\nu\beta}.
\end{eqnarray}

 Similar expressions are deduced for the first derivatives of the metric that are finite over the particle's world lines.

We estimate the finite sum in Eq.~(\ref{E:g1}) by means of an integral, using particle density according to Milne's model (in other words, \emph{we make an statistical average assuming a uniform random process over a surface $\tau$ = constant}). A major simplification comes from the fact that the world lines of the particles are straight lines, for  $ (z^0=\tau , z^i=0)$ is solution of the equations of motion ~(\ref{E:em1}),  and because all the world lines are equivalent due to the Lorentz invariance of the evolution equations,  and thus we only need to calculate the physical metric over a world line of reference $L$ . The integration is taken in the interval $0 < r < r_{max}$. All this greatly simplifies the problem, and we can express  the statistical average of the first order correction to the metric in the form (see the Appendix for details):
\begin{eqnarray}\label{E:staverage}
\langle g_{00}^{(1)}\rangle &=& -\frac{\Omega_{bai}}{4}H_0^2 t^2 g(t,t_i) \ , \ \  \langle g_{0i}^{(1)}\rangle = 0 \\
g(t,t_i)& =& \left(1+\frac{t_i^2}{t^2}\right)^3-4\frac{t_i^2}{t^2}\left(\frac{3}{2}(1+\frac{t_i^2}{t^2})-\frac{t_i}{t}\right),
\end{eqnarray}
and the  mass  $M(t)$ up to the first order as:
 $ M(t)=m\left(1-\frac{1}{2} \langle g^{(1)}_{00}\rangle \right)+C^{(1)}$, where we have built a dimensionless factor $\Omega_{bai}=\frac{8\pi G}{3H_0^2}\mu_{bai}$,  with the baryonic rest mass density at the time of galaxy formation $\mu_{bai}$, and the critical density. Let us point out that the function $g(t,t_i)$ and its first derivative vanishes at $t_i$, and tends rapidly to $1$ when $t$ goes to infinity. \emph{This has the important consequence of producing a positive acceleration at some epoch  $ t > t_i$, as will be seen in section \ref{SS:DEDMv}}.
 Substituting the first order approximation for the metric in Eq.~(\ref{E:M0}) we get, over  the particle of reference $L$, the mass $M(t)$ and the fraction $\frac{dt}{ds} =1-\frac{1}{2} g^{(1)}_{00}$ up to first order in $G$ :
\begin{eqnarray}
M(t) &=& m\left(1+\alpha +\frac{\Omega_{bai}}{8}H_0^2 t^2 g(t,t_i) \right)\label{E:M1}\label{mass}  \\
\frac {dt}{ds}&=&1+\frac{\Omega_{bai}}{8}H_0^2 t^2 g(t,t_i)  \\
M(t)&=& m \frac{dt}{ds} + m \alpha\label{E:factorgamma}
\end{eqnarray}
where we have defined the constant $\alpha$ substituting $ m \alpha$ for  $C^{(1)}$. 

Let us compare our equation $M(t)= m \frac{dt}{ds}+ m \alpha $ with the special relativity formula $E= m \gamma = m \frac{dt}{ds}$. In the latter the $\gamma $ factor is due to the velocity of the particle, and $m( \frac{dt}{ds}-1)$ is the kinetic energy. In our case, the mass $M(t)$ of the reference particle $L$  shown in Figure \ref{Fi:Milne}, has a time increasing  $m ( \frac{dt}{ds} - 1)$  component  formed by contributions from all the particles in its past light cone. We shall refer to $M^{(1)} \equiv m \alpha+m( \frac{dt}{ds} -1)$ as the first order dynamical mass.

Everything we have done for the age of galaxy dominance can be repeated for the two earlier ages of star and free atom dominance. We shall use $(')$ and $('')$ to distinguish the star and the free atom eras. So, the two constants of integration corresponding to these phases will be denoted by $ C'^{(1)}\equiv m' \alpha' $,  $ C''^{(1)}\equiv m'' \alpha'' $, where $m'$ and $m''$ are the baryonic mass of a star and an atom respectively. 

\subsection{\label{SS:3ages}The three ages dominated by particles and the meaning of the constants of integration $C^{(1)}$, $C'^{(1)}$, $C''^{(1)}$}
From equation~(\ref{E:M1}) we get  $M(t_i)= m + C^{(1)}$. This suggests that we can justify a no null value for $C^{(1)}$ by assuming that each galaxy, with baryonic rest mass m, was formed by collapse at some time $t_i$ of other particles, say $N'$ stars, of lesser baryonic rest mass $m'$, verifying $m=N'm'$. By also considering  the stars as a  system of free  gravitating point particles we can  calculate the mass  $M'(t)$ of a star, as we did  for the mass of  a galaxy, to obtain $M'(t)=m' + M'^{(1)}(t)+C'^{(1)}$ for $t'_i < t < t_i$, where $t'_i$ is the epoch of star's formation. Continuity through the surface $\tau = \tau_i$ implies $M(t_i) = N' M'(t_i)= N' (m' +M'^{(1)}(t_i))$, thus we obtain the meaning of $C^{(1)}$, after the substitution $M(t_i)= m + C^{(1)}$: 
  \begin{equation}\label{malpha}
  C^{(1)}\equiv  m \alpha = N' M'^{(1)}(t_i)
    \end{equation}
 The meaning of the new constant $C'^{(1)}$ comes now from the relation $M'(t'_i)=m'+C'^{(1)}$, and implies assuming another phase of free gravitating particles, now the free gravitating atoms. We get a similar result 
 \begin{equation}\label{malphap}
 C'^{(1)}\equiv m' \alpha' = N'' M''^{(1)}(t'_i)
 \end{equation} 
 For times   $ \tau < \tau''_i$ ,  atoms and photons are still entangled, so, we shall assume, as in mechanics of continuous media,  a \emph{principle of local action}, according to which the influence of particles outside the  neighborhood of an atom is erased by  local interactions, in the same way as inside a collapsed system, e.g. the solar system, we do not care about cosmological effects. Therefore, we shall  assume that  the constant $C''^{(1)}=m'' \alpha'' =0$.

 \section{\label{S:MET}The macroscopic energy tensor}
In the two next sections we will make the transition from a discrete system to a continuum  and that always implies to define a kind of averaging of  "microscopic" equations, in our case, the field equations (\ref{E:field}). We have already given  in Eq. (\ref{E:staverage}) the statistical average of the first order metric. The average of  non linear quantities as $\Lambda^{(2)}_{\mu\nu}$ reduces to the product of averages of first order quantities, for our statistical process is a uniform Poisson random process. It remains to define the average of distributional tensor densities of the kind 
 $\sqrt{-g}P^{\mu\nu}=\sum \int p^{\mu\nu}(\lambda)\delta^{(4)}(x^{\rho}-z^{\rho}(\lambda))d\lambda$
with support on  world lines of  particles, parametrized with the Minkowskian time $\tau$ or with the proper time $s$.

Some comments will help to introduce a convenient procedure: 
\begin{itemize}
\item The world lines of the particles are straight lines at  each iteration, as was shown in section \ref{SS:order1}.
\item Up to the first order, the cosmological observer will be of the form $u = \sum_{m=0}^1 u^{(m)}=\sum_{m=0}^1 \gamma^{(m)}(\partial_{t}+(r/t)\partial_{r})$, so the hypersurfaces of constant proper time, corresponding to the auxiliary metric $\eta$ and to  the physical metric  $\sum_{m=0}^1 \langle g^{(m)}\rangle$ will be parallel: the surface $\tau=const.$ coincides with a surface $ s= const.$ ;  but with $\tau > s $, as can be derived  from the expression of $dt/ds$ given above.
\item The two members of the field equation (\ref{E:field}) are  coordinate expressions of  tensorial densities.
\end{itemize}

\begin{figure}[hbt]
\centering\includegraphics{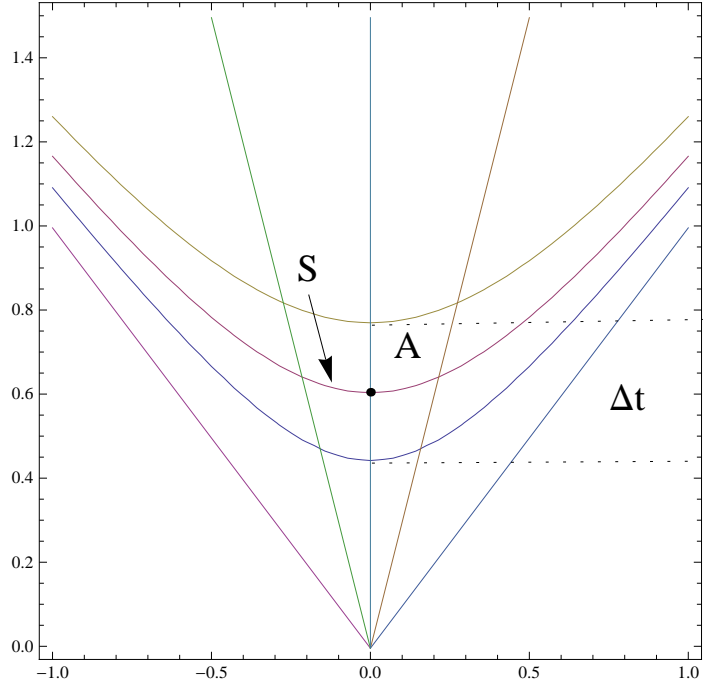}
\caption{Element of volume  $A_x$ centered at a point over the reference line $L$, limited by two surfaces $\tau= const.$  We show  the 3-dimensional surface $S$ necessary to get the density.\label{Fi:volum} }
\end{figure}

Let  $A_x$ be a convenient neighborhood of the point $x$, defined by two neighbor hypersurfaces: $\tau = t+\Delta t/2$ , $\tau = t-\Delta t/2$ (that are also hypersurfaces of constant proper time: $s=s(t) +\Delta s/2 , s= s(t)- \Delta s/2$) and a thin time like cone; and let $S$ be the intersection of the hypersurface $\tau=t$ with $A_x$ as shown in   Figure~\ref{Fi:volum}.

Let $ \varphi_x$ be  the characteristic function  of the set $A_x$:  $ \varphi_x (u)=1$ if $ u \in A_x$ , $\varphi_x (u)=0$ otherwise. 
The average of a distributional tensor density  $\sqrt{-g}P^{\mu\nu}$ is defined as follows
\begin{equation}\label{E:distributional}
\langle \sqrt{-g}\ P^{\mu\nu}\rangle(x)= \lim_{\Delta \lambda \rightarrow 0}\frac{ \left(\sqrt{-g}\ P^{\mu \nu}, \varphi_x \right)}{Vol^{(3)}(S)\Delta \lambda},
\end{equation}
where  the numerator is the action of the distribution on the characteristic function, and $\lambda$ denotes $\tau$ or $s$. This average coincides with the statistical average of the quantities $p^{\mu\nu}$ in the case of a uniform random Poisson process over the surface $\lambda = const.$

Let us give the averages of two suitable examples:
\begin{enumerate}
\item $\sqrt{-g}T^{\mu \nu}_{(s)}=\sum \int m u^{\mu}u^{\nu} \delta^{(4)}(x-z(s))ds$, where the summation is over all the particles, and $u^{\alpha}=\frac{d z^{\alpha}}{d s}$. Then we have
\begin{eqnarray}
\left(\sqrt{-g}T^{\mu \nu}_{(s)}, \varphi_x \right)&=& \sum \int_{s(t)-\Delta s/2}^{s(t)+\Delta s/2} m  u^{\mu} u^{\nu} ds\nonumber\\ 
&& =N(A_x) m  (u^{\mu} u^{\nu})_{s^{*}} \Delta s,
\end{eqnarray}
where  $N(A_x)$ is the number of lines crossing the region $A_x$, and   $s(t)-\Delta s/2 < s^{*} < s(t)+ \Delta s/2$. Using definition (\ref{E:distributional}) one gets the energy tensor for dust:
$\langle \sqrt{-g}T^{\mu\nu}_{(s)}\rangle(x)= n(s) m u^{\mu} u^{\nu}$.

\item $\sqrt{-g}T^{\mu \nu}_{(\tau)}=\sum\int M(\tau)v^{\mu}v^{\nu} \delta^{(4)}(x-z(\tau)) d\tau$, where $v^{\alpha}=\frac{dz^{\alpha}}{d\tau}$. Now we have
\begin{equation}
\left(\sqrt{-g}T^{\mu \nu}_{(\tau)}, \varphi_{x} \right)= \sum \int_{t-\Delta t/2}^{t+\Delta t/2} M(\tau) v^{\mu} v^{\tau}  d\tau\nonumber \\
= N(A_x)M(t^{*}) (v^{\mu} v^{\nu})_{\tau^{*}} \Delta t.
\end{equation}

Using again definition (\ref{E:distributional}) one gets: $ \langle \sqrt{-g}T^{\mu\nu}_{(\tau)}\rangle(x)= n(\tau)M(\tau)v^{\mu}  v^{\nu}$.

\end{enumerate}

 We shall not take any of these as the macroscopic tensor, for what we need is the tensor whose average generates the macroscopic metric tensor. We must consider the averaged field equations
\begin{eqnarray}
 \Box\langle\gamma_{\mu \nu}^{(1)}\rangle &=& -16 \pi G \langle\overline T_{\mu \nu}^{(0)}\rangle \\
 \Box\langle\gamma_{\mu \nu}^{(2)}\rangle +2\langle\Lambda_{\mu \nu}^{(2)}\rangle &=& -16 \pi G \langle\overline T_{\mu \nu}^{(1)}\rangle 
\end{eqnarray}
Here we have taken separate field equations to each order, instead of the form given in Eq.~(\ref{E:field}),  because in our case the world lines are straight lines at all orders (see section  \ref{SS:order1}), thereby there is no implicit dependence on $G$ in the field equation~(\ref{E:field}).

We see that  the macroscopic metric up to order two $\eta_{\mu\nu}+\sum_{m=1}^{2}\langle g_{\mu \nu}^{(m)}\rangle$ is related to the average of the covariant tensor density introduced in section \ref{S:Dyna} that, in the special coordinates we are using, has components  $\sum_{m=0}^{1}\overline T_{\mu \nu}^{(m)}= \sum_{m=0}^{1}(\sqrt{-g}T^{\alpha}_{\mu})^{(m)}\eta_{\alpha \nu} $;  therefore we shall associate   the macroscopic  density energy tensor defined by   $\sum_{m=0}^{1}\langle\overline T_{\mu \nu}^{(m)}\rangle$ to the  discrete system considered here.
Our choice is related, at each step of sequence (\ref{E:sequence}),  to the averaged metric tensor; unlike the two discarded examples.

\subsection{\label{SS:eden}The energy density}

The macroscopic energy density with respect to the macroscopic  $4$-velocity $u$, is  the average $\rho=\langle \overline T_{\mu\nu} u^{\mu}u^{\nu} \rangle$. 
The energy tensor  of each element in the sequence of space-times   is completely determined by the mass series.  Henceforth we  simplify the notation writing $X=\sum_{m=0}^1 X^{(m)} $ for quantities as $\overline T_{\mu\nu}$  or  $u^{\mu}$  up to first order in $G$. The cosmological observer tangent to reference line  $L$ is $u=\partial_t/\sqrt{g_{oo}}$ , so the energy density will be
\begin{equation}
 \overline T_{00}u^0 u^0 =  \sqrt{-g}T^{00}=\sum \int M(\tau) \delta^{(4)}(x-z(\tau)) d\tau,
 \end{equation}
 where  we have used  the definition of our Lorentz covariant tensor, $\overline T_{o o}=\sqrt{-g}T^o_o=g_{o o}\sqrt{-g}T^{o o}$, and shown again that it is a distribution  with support over the word lines.
 
 To get the energy density function at a point $(0,t)$ on the reference line $L$ we consider again the neighborhood  $A_x$   shown  in  Figure~\ref{Fi:volum}. The action of the distribution$ \sqrt{-g}T^{o o}$ on the test function $\varphi_x$   is 
 \begin{eqnarray}
  \left(\sqrt{-g}T^{o o}, \varphi_x \right)&=&\sum \int_{t-\Delta t/2}^{t+\Delta t/2} M(\tau)\varphi_x (z(\tau))d\tau\nonumber\\
  &&=  N(A_x) M(t^{*})\Delta t,
 \end{eqnarray} 
where $N(A_x)$ is the number of lines crossing the region $A_x$, and $t-\Delta t/2 < t^{*} < t + \Delta t/2 $. The macroscopic energy density, $\rho$ , is the average
  $\langle \sqrt{-g}T^{00}\rangle$, defined in equation (\ref{E:distributional}) of this section,  that is
 \begin{equation}\label{density}
 \rho=  \lim_{\Delta t \rightarrow 0} \frac{\left(\sqrt{-g}T^{o o}, \varphi_x \right)}{ Vol^{(3)}(S)\Delta t} = n(t) M(t),
\end{equation} 
where $n(t)$ is the number density of particles. By substituting the mass up to the first order we get
\begin{equation}
\rho= n(t) m\left(1+\alpha +\frac{\Omega_{bai}}{8}H_0^2 t^2 g(t,t_i) \right) 
\end{equation}

\section{\label{S:RW}RW  models corresponding to each order of approximation}

If it were possible to  solve the field equations to each order, and  prove the convergence of the series giving the metric and the energy tensor in the coordinate system $ ( t , x^i)$, then, we know that a change of coordinates $ t = t(s, \overline x^i) , x^i= x^i( s , \overline x^i) $  would put the space time 
$(\sf M , \eta+\sum_{m=1}^{\infty}\langle g^{(m)}\rangle , \sum_{m=0}^{\infty} \langle\overline T^{(m)}\rangle)$ into a RW universe in comoving coordinates. However, it is not so complicated, because a RW model is completely determined by giving the curvature index $k$ (we shall take $k=0$),  the energy density as function of the scale factor, and solving the equations
 \begin{eqnarray}
  \rho(a) &= & \frac{3H_0^2}{8\pi G}\left(\frac{\Omega_m}{a^3}+f(a)\right)  \\ 
  \frac{da}{ds} &=& H_{0}\sqrt{\frac{\Omega_{m}}{a}+a^{2}f(a)}. 
  \end{eqnarray}
 We have obtained  the energy density up to the first order in $ G$, at any point of the reference line $L$, as function of the time coordinate $t$. So, we could associate  a RW model  to each order of approximation simply by substituting  $n(a)=n_{0} /a^{3}$ for $n(t)$ in our energy density and finding  the function $t(a)$.  The true model is the corresponding to the limit of the series, but it happens that, \emph{and we do not know why}, the model associated to the first order energy tensor is an excellent approximation to the $\Lambda CDM$  concordance model.

Next we shall obtain the pair $(\Omega_m , f(a))$ corresponding to the first elements of the sequence.

 \subsection{RW model corresponding to the order zero}  At zero order, $\rho=0$,  then $ \Omega_m =0$ and $f(a)=0$  and we get the Milne universe.
 \subsection{\label{SS:first} RW model corresponding to the first order}To the  first order in the metric corresponds the zero order in the mass $M^{(0)}=m$ and $ \rho= n(a) m  $, as follows from Eq.~(\ref{density}). Therefore, for the first order we get  $ \Omega_m = \Omega_{ba}$ , $f(a)=0$, i.e., we have dust as energy tensor but only in the first approximation to the metric, which is a good approximation only for times near to the particle formation.  Recall that in general it is accepted  an Einstein de Sitter universe as the macroscopic description of a universe made of collision-less particles, see the end of section \ref{SS:eden}.

 \subsection{RW model corresponding to the second order} We have got an expression for the energy density  
 \begin{equation}
     \rho=n(a) m\left(1+\alpha+\frac{\Omega_{bai}}{8}H_0^2 t^2 g(t,t_i) \right)
 \end{equation}    
 but  it is incomplete for we still need the function $t(a)$  expressing  the coordinate $t$ as a function of the expansion factor.  It should be  the solution of the equation $da/dt = (da/ds)(ds/dt)$
\begin{equation}
    \frac{da}{dt}=\frac{ H_0\Omega_m^{1/2}}{a^{1/2}}\left(1+O(\Omega_{ba})\right).
\end{equation}
 We shall only consider the dominant contribution, ignoring terms of first order in  $\Omega_{ba}$
\begin{equation}
H_0 t = \frac{2a^{3/2}}{3\Omega_m^{1/2}}(1+O(\Omega_{ba}))
\end{equation}
The function $\rho(a)$ up to the first order in G is then
\begin{equation}
\rho=n(a) m\left(1+\alpha+\frac{\Omega_{bai} a^3}{18\Omega_m} g(a,a_i) \right) 
\end{equation}
\begin{equation}
g(a,a_i) =  \left(1+\frac{a_i^3}{a^3}\right)^3-4\frac{a_i^3}{a^3}\left(\frac{3}{2}\left(1+\frac{a_i^3}{a^3}\right)-\frac{a_i^{3/2}}{a^{3/2}}\right),
\end{equation}
where  $g(a,a_i)$ comes from $g(t,t_i)$ after substituting $t=\frac{2a^{3/2}}{3H_0\sqrt{\Omega_m}}$. The function $g(a,a_i)$ vanishes at $a=a_i$ and tends rapidly to unity.

 By writing  the energy density in the form
\begin{equation}
\rho(a)=   \frac{3H_0^2}{8\pi G}  \left(\frac{\Omega_{ba}(1+\alpha)}{a^3} + f(a)\right)
\end{equation}
the pair $(\Omega_m , f(a))$ comes in directly 
\begin{eqnarray}
\Omega_m & = & \Omega_{ba}(1+\alpha)  \\
f(a) & = & \frac{\Omega_{ba}}{18a_i^3(1+\alpha)}g(a,a_i)\label{E:f(a)}.
\end{eqnarray}
Standard calculations produce the pressure and the acceleration of the model
\begin{eqnarray}
 p(a)  & = &    \frac{3H_0^2}{8\pi G}\left(-f(a)+\frac{1}{3}a \frac{df}{da}\right)    \\
\frac{\ddot a}{a} & = &  - H_0^2 \left(\frac{\Omega_m}{2a^3}-f(a) +\frac{1}{2}a\frac{df}{da} \right)
\end{eqnarray}

\section{Dark energy and dark matter phenomena}

The mass $M$ appearing in the energy tensor of point-like gravitating particles is not a constant. Under the conditions of cosmology, \emph{local isotropy at any point}, it turns out to be  a monotone increasing function of time. This affects the  energy density of a universe made of gravitating particles in a way that could could make the existence of new forms of dark matter and dark energy unnecessary.

\subsection{\label{SS:DEDMv}Dark energy and matter values,  $\Omega_{\Lambda}, \Omega_{dm}$, in the galactic phase, and the redshift $z_i$ of formation of galaxies} 
\begin{figure}[hbt]
\centering\includegraphics{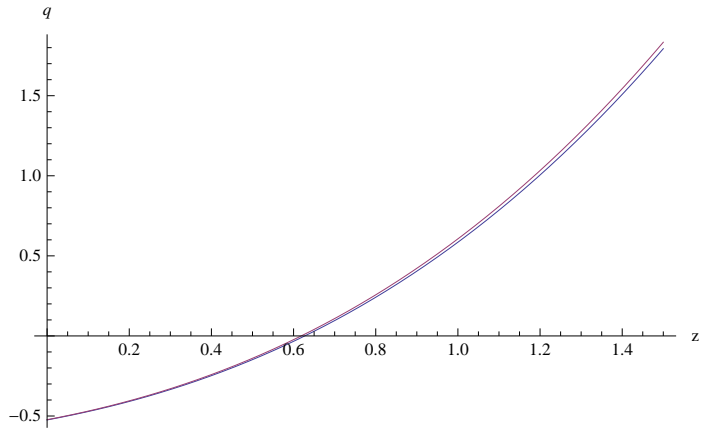}
\caption{We show predictions for deceleration factor, corresponding to our model (lower curve) and the concordance model.}\label{Fi:q(z)}.
\end{figure}

In our model the dark values are related to the galaxy formation epoch
It is amazing that the first order energy tensor  be  enough to obtain an energy component $f(a)$ which reproduces the main predictions of the concordance model for the galactic epoch.  We have found that  $g(a,a_i)$ tends asymptotically to the unity, and  f(a)  to  a limit value.  It's convenient to express $a$ as function of redshift: $a=1/(1+z)$. The function $f(a(z))$ will be referred to as $f(z)$. We can estimate  the redshift  of  galaxy formation $z_i$ by requiring that, at the present epoch, the component $f(z)$ coincides with the observed cosmological constant $f(0)=\Omega_\Lambda=0.683$. In terms of redshift, this amounts, using Eq.(\ref{E:f(a)}), to the equation
\begin{equation}
(1+z_i)^3 g(0,z_i)= 18 \Omega_{\Lambda}\frac{\Omega_{ba}(1+\alpha)}{\Omega_{ba}^2}\ , 
\end{equation}
and, assuming that all what has been observed as dark matter can be explained as the dynamical mass introduced in this paper, we identify   $\Omega_{ba} \alpha = \Omega_{dm} =0.268 $, and  solving for the redshift we get  the value $z_i =10.76$, which seems reasonable. This determine completely the energy density for the galactic epoch. The point is that the predictions of this model are equivalent to those of the concordance model as far as: 
\begin{enumerate}
\item The energy density  $f(z)$ tends to a constant that coincides with the cosmological constant if galaxies were formed at redshift $z_i \approx 11$.
\begin{figure}[hbt]
\centering\includegraphics{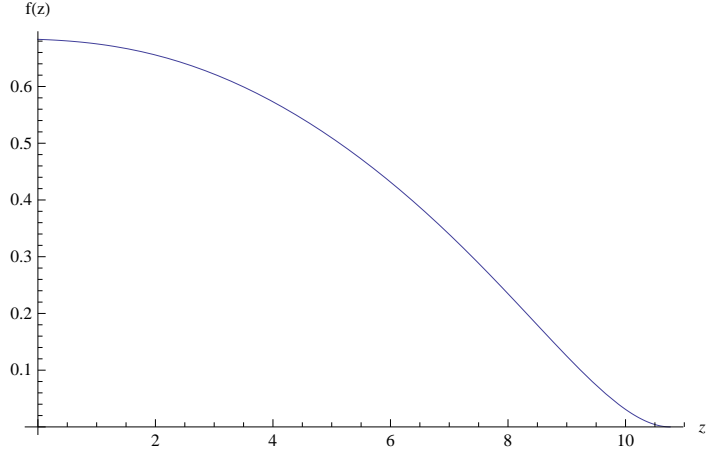}
\caption{"Dark energy"  dependence on redshift for the galactic age. It tends to a limit value that we identify with the observed value for $\Omega_{\Lambda}. $}\label{Fi:f(a)}
\end{figure}

\item The distance moduli  ($\mu=5 \log_{10} d_l +25$)-redshift relation obtained with the energy component $f(z)$ agrees, with great precision, with the predicted by the concordance model.
\begin{figure}[hbt]
\centering\includegraphics{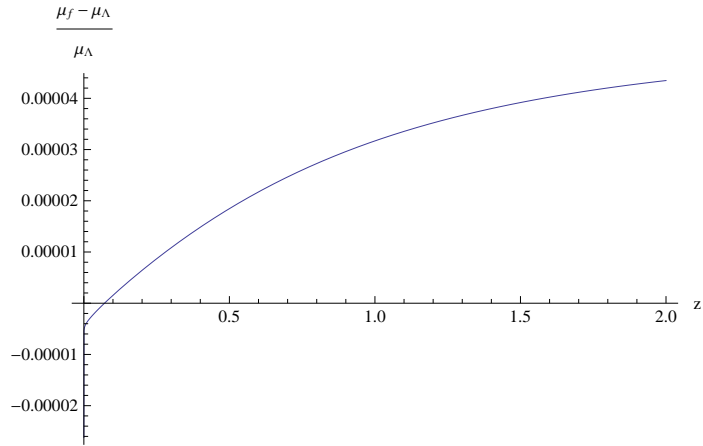}
\caption{We show the tiny discrepancy between distance moduli prediction by the concordance model $\mu_{\Lambda}$ and our prediction $\mu_f$.}\label{Fi:deltamu}
\end{figure}

\item Recent positive acceleration. We obtain positive acceleration at redshifts $ z < 0.625$ as in the concordance $\Lambda CDM$ model, if galaxies surged at redshift  $z_i \approx 11$. In our model a positive acceleration is a recent phenomena because it is a fact  linked to the formation of galaxies. The deceleration parameter $q(z)$ as compared with the cosmological constant prediction is shown in  Figure~\ref{Fi:q(z)}.

\end{enumerate}

\subsection{\label{SS:DM}Dark matter  value, $\Omega'_{dm}$,  in the stellar phase,  and the redshift, $z'_i$, of formation of the stars }

We consider  the era of the stars, $ z'_i < z <z_i$, also dominated by particles  expanding  in a local  isotropic way everywhere, so, placing  a $( ' $)  over quantities referred to this  phase  we can write $\rho'(z)=   \frac{3H_0^2}{8\pi G}  \left(\Omega_{ba}(1+\alpha')(1+z)^3 + f'(z)\right)$ , with $f'(z)  =  \frac{\Omega_{ba}}{18(1+\alpha')}(1+z'_i)^3 g(z,z'_i)$ , for its energy density.
We shall estimate the redshift of  star formation, following three steps:
\begin{enumerate}
 \item First we  impose \emph{continuity of energy density at the galaxy formation epoch} $z_i$ : $\Omega_{ba} \alpha' +\frac{ f'(z_i)}{(1+z_i)^3}=\Omega_{ba}\alpha$, and obtain the equation
\begin{equation}\label{E:alphap}
(\alpha')^2+(1-\alpha)\alpha'-\alpha+\frac{1}{18}\left(\frac{1+z'_i}{1+z_i}\right)^3 g(z_i,z'_i)=0.
\end{equation}
\item Like the galaxies, the stars are the result of the collapse of $N''$ particles of smaller mass $m''$, say atoms; and as above, we shall now place $('' )$ over quantities referring to the atomic phase. Recombination occurs at $z\approx1100$, but decoupling of matter from radiation starts at $z=300$, and an effective decoupling epoch may be approximated  by $z\approx 100$, \cite{Scott}. Then, we shall assume that the free-atom era,  with redshifts $z'_i < z < z''_i = 100$, is also well described as a system of free gravitating particles; and  requiring  \emph{continuity of energy density at the star  formation epoch} $z'_i$ : 
$\Omega_{ba} \alpha'' +\frac{ f''(z'_i)}{(1+z'^3_i)}=\Omega_{ba}\alpha'$, we get the equation

\begin{equation}\label{SS:p0}
(\alpha'')^2+(1-\alpha')\alpha''-\alpha'+\frac{1}{18}\left(\frac{1+z''_i}{1+z'_i}\right)^3 g(z'_i,z''_i)=0.
\end{equation}

\item For redshifts  $ z < z''_i$,  atoms and photons are still entangled, in section \ref{SS:3ages}, we accepted  a \emph{principle of local action} to justify $C''^{(1)}=m'' \alpha''=0$, which means that the increase in atomic mass, due to the gravitational interaction with all the atoms in the past light cone, is zero. Substituting $\alpha'' =0$ in the last equation one gets $\alpha'$ as a function of the redshifts $ z'_i$ and $z''_i$
\begin{equation}\label{SS:p1}
\alpha'(z'_i,z''_i)=\frac{1}{18}\left(\frac{1+z''_i}{1+z'_i}\right)^3 g(z'_i,z''_i)
\end{equation}
and then, taking into account Eq.~(\ref{E:alphap}), we get the equation
\begin{equation}\label{SS:p2}
\alpha'(z'_i,z''_i)^2+(1-\alpha)\alpha'(z'_i,z''_i)-\alpha+\frac{1}{18}\left(\frac{1+z'_i}{1+z_i}\right)^3 g(z_i,z'_i)=0
\end{equation}
that after  substituting  the values $z''_i = 100$ ,  $\alpha = (1-\Omega_{\Lambda}-\Omega_{ba})\Omega_{ba}^{-1}$ and $z_i=10.76$,  we can solve 
to determine the redshift of star formation at $z'_i = 20.7$
(let us point out that we would had  obtained bigger values for the star formation redshift if we had taken $z''_i$ in the interval $ 100 < z''_i < 300$; so, for $z''_i=200$ we obtain $ z'_i = 43.4$).  Finally, substituting $(z'_i , z''_i)=(20.7 ,  100)$ into Eq.~(\ref{SS:p1})we get the dark matter component for the star's phase: $\Omega_{ba} \alpha' = 0.267$.
\end{enumerate}

\subsection{Dark energy and dark matter have a common root}

Dark energy and dark matter show up as two consequences of a sort of \emph{Mach's principle}: 
 Although at zero order in  Einstein's evolution each star which will form the galaxy has initially  a definite rest mass $M'^{(0)}=m' $ (an intrinsic property of the  particle, that  we identified by its baryonic mass), we have remembered  Mach because  the first order dynamical mass $M'^{(1)}$ increases with time due to the influence of more and more stars over the past light cone. The mass aggregated by this process from $ z'_i = 20.7$ until  $z_i = 10.76$ , i.e. $N' M'^{(1)}(z_i) $, plus the mass  obtained in the free-atom-star transition $N'N''M''^{(1)}(z''_i) $, could be the so called galaxy's  dark matter at its birth, i.e. $C^{(1)}\equiv m \alpha$.

  Once the galaxies appear at $z=z_i$, their gravitational interaction begins, and  the previous process starts again but now with a galaxy as the new particle. At zero order the mass of the galaxy is its baryonic mass $m=N' m'$, and  the first order dynamical mass $M^{(1)}(z)$ for $z < z_i$ produces the same effects as  the so called dark energy. Multiplying $M^{(1)}(z)$ by the number density and dividing  by the critical density  one gets the  function $f(z)$, whose main observational effect is an acceleration of the universe and  supernova Hubble diagrams undistinguishable from those predicted by the concordance model.  At very large scales, in the galactic phase, dark matter is described by the summand $\Omega_{ba} \alpha$ , and dark energy by the second sumand $f(z)$ in the expression for  energy density.

\section{ \label{S:dmg}Dark matter phenomenon at galactic scales}

Up to now our model has considered galaxies as point particles. To consider them as inhomogeneities contradicts the local isotropy, on which we have based our discussion. However, we can get  reasonable results  by stating a hypothesis about how to use the density formula $\rho(r , z)=\mu_{ba}(r)(1+ \alpha +\frac{\Omega_{ba}(1+z_i)^3}{18(1+\alpha)}g(z , z_i))$ inside a local inhomogenity.  We assume  that the mass of a star at the center of the galaxy will not be affected by the cosmological increase in mass given in Eqs. (\ref{mass}), (\ref{mstar}), but it must be considered in some proportion at great  distance from the center because then we are improving the local isotropy conditions. To make this more explicit we shall use  a  rough newtonian description of the galaxy as a Plummer's model, which is analytical  and has finite total mass. So, we consider a baryonic density 
$\mu_{ba}=\frac{3M}{4 \pi R^3}(1+\frac{r^2}{R^2})^{-5/2}$ which produces a newtonian potential $\Phi_{ba}=-\frac{GM}{R}(1+r^2/R^2)^{-1/2}$. Now, the closer the gravitational potential goes to zero the more we sense the effect of increasing mass, for we are recovering the homogeneous universe. Let us define the function $\chi (r) =1- \frac{\Phi_{ba}(r)}{\Phi_{ba}(0)}$ which vanishes at the center and tends to unity at infinity, and assume that a Plummer's galaxy, immersed in a universe should have  mass density

\begin{equation}
\rho(r , z)=\mu_{ba}(r)+\chi (r) \mu_{ba} (r) (\alpha +\frac{\Omega_{ba}(1+z_i)^3}{18(1+\alpha)}g(z , z_i)) .
\end{equation}
The  density of dark matter at distance $r$ will then be $\rho_{dm}(r)= \rho(r,z)-\mu_{ba} (r) $, despite the fact that the galaxy is made only of baryons. Let us confront  some predictions of this  coarse model:\begin{enumerate}
\item We can estimate the  " dark mass "  inside a sphere of radius $r$ as 
\begin{equation}
 M_{dm}(r , z)= \int_0^r   (\rho(r , z) -\mu_{ba} (r)) 4 \pi u^2 du 
 \end{equation}
\item The fraction of  " dark mass " inside a sphere of radius $r$ is
\begin{equation}
f_{dm}(r , z)=\frac{M_{dm}(r , z)}{M_{ba}(r)+M_{dm}(r , z)} \ ,
\end{equation}
\end{enumerate}
with recent observations. 

Baryons and dark matter has been recently disentangled \cite{Suyu}  in the spiral lens B1933+503, observed at redshift $z=0.755$. We have chosen a Plummer model with $M=2.6 \times 10^{11}M_{\odot }$ , and scale factor $R=12 Kpc $ in order to get the intersection of  baryonic and dark matter profiles at $ r= 10 Kpc$. Our coarse model of the galaxy agrees qualitatively with the observations inside the crossing point, and also quantitatively in the outer region. 
\begin{figure}[hbt]
\centering\includegraphics{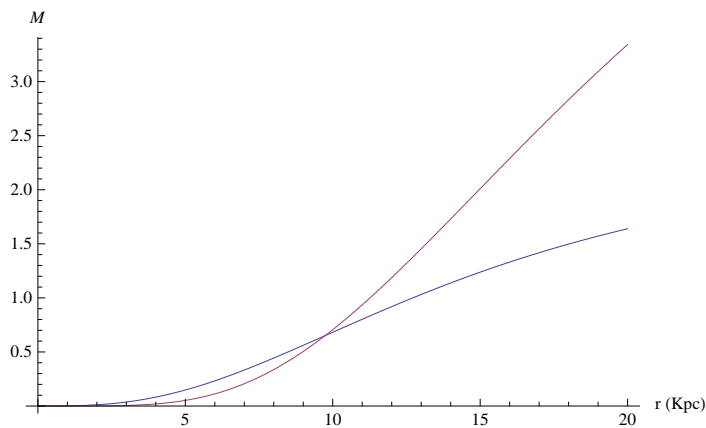}
\caption{Mass inside radius $r$ for baryonic and dark matter components in the lens $B1933+503$, using a Plummer's model. The straight portion of the dark component is liable of the constant rotational velocity of hydrogen atoms.}\label{Fi:M(r)}
\end{figure}

\begin{figure}[hbt]
\centering\includegraphics{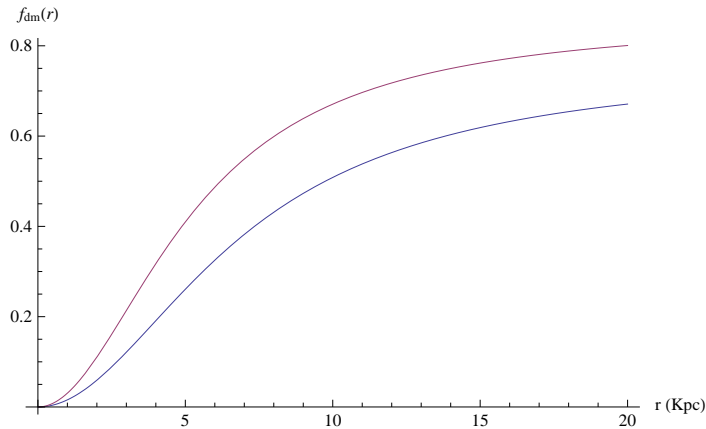}
\caption{Dark matter fraction of the same galaxy at redshifts $z=0.755$, lower curve, and $z=0.138$, upper curve.}\label{Fi:fractiondm}
\end{figure}

As for the dark matter fraction profile, we get good values for $r > 10 Kpc$. We have also  shown in Figure~\ref{Fi:fractiondm} the dark matter fraction this galaxy would have at less redshift, say $z=0.138$.  In the outer regions it reaches the value $f_{dm}=0.8$, and
this value also agrees with observations \cite{Dutton}  of dark matter fractions in the lens $SDSS J2141-0001$  at redshift $z=0.138$.

\begin{figure}[hbt]
\centering\includegraphics{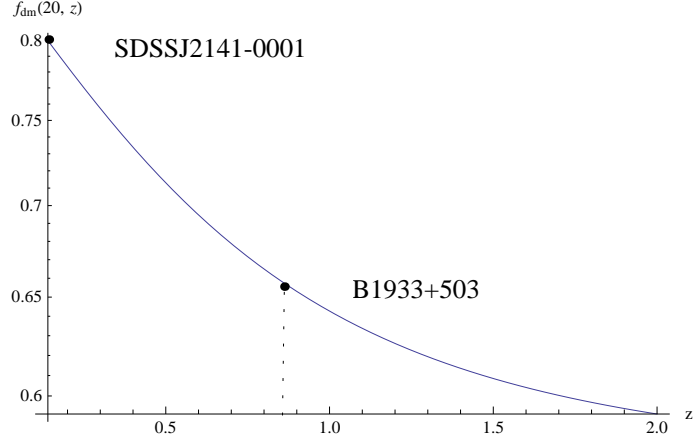}
\caption{Evolution of dark matter fraction, at $20 Kpc$ from the galactic center, showing two recent observations.}\label{Fi:evolution}
\end{figure}

 The dependence on redshift of the dark matter fraction at a fixed radius, $r = 20 Kpc$  is shown in  Figure~\ref{Fi:evolution}. It should decrease with redshift because what we call dark matter is mass added to the initial rest mass due to the gravitational interaction with particles over its past light cone. 

\section{Summary and conclusions}
This paper aims  to find the macroscopic energy density which corresponds to a homogeneous and isotropic universe dominated by free gravitating particles. 
We have revised the statement that it should be the energy density of  dust. As discussed in section \ref{SS:eden}, one arrives to this conclusion by averaging the energy tensor for point like particles  $\langle T^{\rho \sigma}_{(s)}u_{\rho}u_{\sigma}\rangle$ = $n m$. Then, Einstein's equations produce an Einstein de Sitter space time.

Nevertheless, inverting the process, that is to say, obtaining  first the metric $g^{(1)}_{\mu\nu}$ by solving the the first order field equations using the zero order energy tensor $\overline T^{(0)}_{\mu\nu}$ of section \ref{S:Dyna}, and after that,  averaging the first order energy tensor $\overline T^{(1)}_{\mu\nu}$ obtained with this metric,   we get an energy density  very similar to the one of the  $\Lambda CDM$ cosmological model: 
\begin{equation}
T^{(0)}\rightarrow  g^{(1)}_{\mu\nu}  \rightarrow  \sum_{m=0}^{1}\langle \overline T^{(m)}_{\mu\nu}u^{\mu}u^{\nu} \rangle  = \rho= n(a)\sum_{m=0}^{1}M^{(m)}(a) \approx  \rho_{\Lambda CDM}
\end{equation}

This result prompts us to represent the mass of a particle as  $\sum_{m=0}^{\infty}M^{(m)}(a)$. The zero order is the mass that it  would have if it were an isolated particle, we have identified it as its baryonic mass; the higher orders represent the contribution of the interaction with the rest of particles, we refer to them as  dynamical mass (see the comment at the end of section \ref{SS:order1}). 

We have focused on  the ages in which, apparently, the universe was dominated by  free gravitating particles, namely, the epoch of the free atoms:  $ z''_i \approx 100 < z < z'_i \approx 21$, the epoch of   stars: $ z'_i \approx 21 < z < z_i \approx 11$ and the epoch of galaxies:  $ z_i \approx 11 < z < 0$.
As we have said, In each phase the mass of a particle at zero order is its baryonic mass, i.e., $M^{''(0)}=m''$ for an  atom, $M^{'(0)}=m'=N''m''$ for a star and  $M^{(0)}=m=N'm'$ for a galaxy. 
 
 The mass up to the first order of a particle of the universe  is an increasing function of the scale factor, because it  is the sum of contributions from  all the particles intersecting  its past light cone, and   the number of these  increases with time:
\begin{eqnarray}
m''+M''^{(1)}(a) & = & m''+ \frac{m''  a^{3}}{18 a_{i}^{''3}} g(a,a''_i) \ \ , \  for \  free \ atoms   \\
m'+M'^{(1)}(a) & = &m'+ m' \alpha'+ \frac{m'  a^{3}}{18(1+\alpha') a_{i}^{'3}} g(a,a'_i) \ \ , \  for \  stars \label{mstar}  \\
m+M^{(1)}(a) & = & m+  m \alpha + \frac{m  a^{3}}{18(1+\alpha) a_{i}^{3}} g(a,a_i)   \ \ , \ for \ galaxies,
\end{eqnarray}
where $ m \alpha =N' M'^{(1)}(a_i) $   refers to the dynamical mass gained by the stars that collapsed to form a galaxy,  i.e., their content of "dark matter" at the moment of galaxy formation, and  $ m' \alpha' =N'' M''^{(1)}(a'_i)$  refers to the dynamical mass gained by the atoms that collapsed to form a star, as can be derived from equations (\ref{malpha}), (\ref{malphap}).

\subsection{Dark energy and dark matter at large scales}
 Henceforth we shall use  the redshift instead of the scale factor as independent variable. Multiplying the masses  $M''^{(1)}(z)$,  $M'^{(1)}(z)$,  $M^{(1)}(z)$ by the number density corresponding to each phase  we have obtained the energy density:
\begin{eqnarray}
 \rho''(z) & = &   \frac{3H_o^2}{8\pi G}  \left(\Omega_{ba}(1+z)^3 +  \frac{\Omega_{ba} (1+z''_i)^3 }{18}g(z , z''_i)\right),  for \   atoms \\
 \rho'(z) & = &   \frac{3H_o^2}{8\pi G}  \left(\Omega_{ba}(1+\alpha')(1+z)^3 +  \frac{\Omega_{ba}(1+z'_i)^3}{18(1+\alpha')}  g(z , z'_i)\right), stars   \\
 \rho(z) & = &  \frac{3H_o^2}{8\pi G}  \left(\Omega_{ba}(1+\alpha)(1+z)^3 +  \frac{\Omega_{ba}(1+z_i)^3 }{18(1+\alpha)} g(z,z_i)\right), galaxies\label{Su:rho}
\end{eqnarray}
These expressions allow us to identify, in each phase, what plays the roles of dark matter and dark energy, e.g., in the galactic age, the "dark matter" and "dark energy" components  will be $\Omega_{ba}\alpha$ and $f(z)= \frac{\Omega_{ba}}{18(1+\alpha)}(1+z_i)^3  g(z,z_i)$ respectively.

The parameters of our model are the baryonic density $ \Omega_{ba}$ and the redshifts $ z''_i , z'_i , z_i $, corresponding to the effective decoupling of atoms and radiation, the formation of stars and galaxies respectively. Requiring to the energy density continuity through the interface surfaces,  we get relations between the redshifts of stars and galaxies formation and the amount of acquired dynamical mass ('' dark matter '').

 \subsubsection{The meaning of the dark energy}
 The second summand on the right hand of the energy density for each phase: ($ f''(z), f'(z) , f(z)$ ) tends rapidly to a limit $(\frac{\Omega_{ba}(1+z''_i)^3}{18}, \frac{\Omega_{ba}(1+z'_i)^3}{18(1+\alpha')},  \frac{\Omega_{ba}(1+z_i)^3}{18(1+\alpha)})$. So, in our model, this component of the energy density will manifest itself as a cosmological constant when the universe approaches the end of each phase.
 
 Our model gives very good results for the galactic age:
 \begin{enumerate}
\item  By identifying  $f(0)=\Omega_{\Lambda}=0.683$ , and $\Omega_{ba}\alpha =\Omega_{dm}=0.268$    we have obtained the redshift of formation of the galaxies $z_i=10.76$, which  seems a reasonable result.
\item The distance moduli-redshift diagram is practically identical to that predicted by the concordance $\Lambda CDM$ model, as shown in Figure \ref{Fi:deltamu}.
\item  The acceleration of the Universe (see Figure \ref{Fi:q(z)}) is a recent phenomenon, for $ z<0.625 $, because it is linked to the birth of the galaxies.
\end{enumerate}

 \subsubsection{The meaning of the dark matter}
 The summand $\Omega_{ba} \alpha$ in the second term of equation (\ref{Su:rho})  is indistinguishable from the dark matter parameter in the $\Lambda CDM$ model, hence we have identified $\Omega_{ba} \alpha =\Omega_{dm}= 0.268$. In our approach the meaning of this figure is $\Omega_{ba} \alpha =\frac{3H_o^2}{8\pi G}n(0) m \alpha$,  where  $n(0)$ is the  number density of galaxies at redshift zero, and  from equations (\ref{malpha}), (\ref{malphap}), and recalling that $m=N'm'$ we can write
 \begin{eqnarray}
 \Omega_{ba} \alpha& =& \frac{8\pi G}{3H_o^2}n(0)N'M'^{(1)}(z_i) \\
 \Omega_{ba} \alpha' &= &\frac{8\pi G}{3H_o^2}n(0)N' N'' M''^{(1)}(z'_i)
 \end{eqnarray}
, where  $N'$ is the number of stars  that formed a galaxy, $N''$ the number of atoms that formed a star,  $M'^{(1)}(z_i)$ is  the dynamical correction to the star's  rest   mass at the galactic formation epoch, and  $M''^{(1)}(z'_i)$ the correction to the atomic rest mass mass at the star formation.
 
 Furthermore, in section ~\ref{SS:DM}, requiring the energy density to be continuous through the interfaces $ z=z_i \ , z=z'_i $ we have obtained two more results:
 \begin{enumerate}
  \item  Using equation~(\ref{SS:p1})
obtained by continuity of energy density through the surface $ z= z'_i$ of star's formation, 
  we have got, taking  $z''_i=100$,  the dark matter component  during the star's phase as     $ \Omega_{ba}\alpha' = 0.267 $,  very similar to the dark component of the galactic phase $\Omega_{ba}\alpha= 0.268$. However, as discussed at the end of Sec. ~(\ref{SS:DM}), this estimation depends on which value of $z''_i$ is taken from the interval $(100,300)$.
 \item  Using equation~(\ref{SS:p2})
obtained by continuity of the energy density through the surface $ z= z_i=10.76$ of galaxy formation, and assuming  $z''_i=100$, we have  got  the redshift of star formation $z'_i =  20.7$. Although had we taken for  $z''_i$ the mean value of the interval $(100,300)$ we had got  the  birth of the stars at  $z'_i =  43.4$.

\end{enumerate}

Let us make a necessary remark:   we have not proven that dark matter and dark energy do not exist in the form of some exotic components, we have assumed that they do not exist. In fact, if, for example, we were forced to delay the dominance of the galactic component from redshift of order $11$ to redshift of order $3$, our model would need some quantity of exotic matter and energy.

\subsection{Dark matter at galactic scales}
Even though in this paper galaxies are treated as point particles, in section  \ref{S:dmg} we showed how to get some insights on the dark matter fraction profile. Using a newtonian description of the inhomogeneity and taking its newtonian potential as an index of accomplishment of the conditions of homogeneity we have suggested that:

\begin{enumerate}
\item The mass of a star in a galaxy is of the form 
\begin{eqnarray}
&&m'(r , z) = m' +m'  \chi (r) \left(  \alpha +\frac{\Omega_{ba}(1+z_i)^3}{18(1+\alpha)}g(z , z_i)\right) \\
&&\chi (r)  = 1- \frac{\Phi_{ba}(r)}{\Phi_{ba}(0)}
\end{eqnarray}

\item Multiplying by the number density of stars $n'(r)$ one obtains the energy density inside the galaxy $ \rho^{gal}(r,z)= n'(r)m'(r,z)$. We identify the dark matter inside a galaxy as the difference $\rho'_{dm}(r,z)=\rho'(r,z)-n'(r)m'$, that is to say
\begin{equation}
\rho^{gal}_{dm}(r,z)=\mu_{ba}^{gal}(r)\chi (r)  \left(\alpha +\frac{\Omega_{ba}(1+z_i)^3}{18(1+\alpha)}g(z , z_i)\right)
\end{equation}
\end{enumerate}
To illustrate the theory we have considered a Newtonian Plummer's model to describe the distribution of baryonic matter (stars) in a galaxy.  Notwithstanding the inaccuracy of a Plummer's model to describe a spiral galaxy, we have  reproduced the dark matter fraction profiles  recently observed, using the lens effect, in the spiral galaxy $B1933+503$ (see Figure~\ref{Fi:M(r)}); and we have predicted an evolution of the dark matter fraction, as shown in figures~\ref{Fi:fractiondm} and \ref{Fi:evolution}. The last figure is  compatible with a couple of recent observations.
 
\vspace{0.2cm}

 \subsection{Final comments: First order effects, backreaction and black hole universes.}

\begin{enumerate}
\item  It is amazing that a cosmological model based on the first iteration of the relativistic $N$ body problem, could reproduce so accurately, as it does, the dark phenomena. There should be some reason why non linear effects must be negligible. Meanwhile,  this work is only  a new insight, among other ones, to the dark problems. 
\item
Many works have been  published  on cosmological backreaction from small scale inhomogeneities. 
The averaging of the nonlinear terms in Einstein's equations does not commute in general with evaluating them with the averaged metric. Therefore, one can need to add a term  to the energy tensor in order that the averaged metric satisfy the field equations.  Though it is not free of controversy this is up to day the main idea  to explain the dark phenomena without exotic physics ( see \cite{A1}  for references).

The results of this paper are not related with backreaction from small scale inhomogeneities, even more,  this problem cannot be studied  within the settings of the paper. This is because one considers a uniform Poisson process of particles over space like hipersurfaces $\Sigma_{\tau}$(see Fig A1). The metric components obtained from the $N$-body  problem depend on the particle positions. The average is a statistical average, that is,  the mean value respect to the joint distribution probability. Then  there is no problem with the nonlinear terms  when averaging the field equations at each order, see equations (23-24) of Sec 4,
because the average of the quadratic terms $\Lambda_{\mu \nu}^{(2)} $ equates the product of the average of the factors. This is easy to understand. The first order metric components are functions of the form $g=\sum_k \xi(x, \bar x_k)$, where $\{\bar x_k \}, k= 1, 2 ,3 ...A$, is a realization of a Poisson process with density $ n^{(0)}(r,t)=Nt(t^2-r^2)^{-2}$,  on a ball of radius $r_{mx}$ contained in the hipersurfaces $\Sigma_t$ (see Fig A1). Any quadratic term will be of the form $ g  h = \sum_{ks} \xi(x, \bar x_k) \eta(x, \bar x_s)$, and the statistical average will be the product of two integrals $\langle  g h\rangle = \langle g  \rangle\langle  h \rangle$, with
\begin{equation}
\langle g  \rangle = 4\pi \int_0^{r_{mx}}  n^{(0)}(r,t) \xi(x, \bar x) r^2 dr \ , \  \langle h  \rangle = 4\pi \int_0^{r_{mx}}  n^{(0)}(r,t) \eta(x, \bar x) r^2 dr
\end{equation}
and $r=\mid \bar x \mid $, because the joint probability density is 
\begin{equation}
 f(\bar x_k , \bar x_s) = A^{-2} n^{(0)}(r_k,t)n^{(0)}(r_s,t).
 \end{equation}
  Hence, our results  have no relation with the back-reaction. Nevertheless, 
the average of  the energy tensor $\overline T_{\mu \nu}$ that corresponds to the second order metric, which depends on the metric calculated at first order (see Eq. 28), does not coincides with the average of the tensor $T_{\mu \nu}$, which does not depend on the metric. So in a certain sense this no coincidence could be interpreted as the backreaction of the metric (gravitational interaction between the particles) on the energy tensor.

\item 
 A recent series of studies developes the idea of 
 \emph{black hole universes}, with roots in a paper by Lindquist and Wheeler \cite{2}. Using the  $3+1$  form of the Einstein equations they obtain solutions to the constraint equations that describe blackhole singularities, at the center of cubes with periodic boundary conditions at the opposite faces \cite{3}, or distributed on a $ S^3 $ hipersphere \cite{4}, that simulate a discrete uniform universe. They 
 are  arriving  to the conclusion that the expansion law of a \emph{black hole universe} approach to that of the pressure-less EdS universe when the number of BH inside the Hubble radius increases. At first sight this result seems to contradict this paper.
  But 
  the evolution equations of the \emph{black hole universe} takes  a null energy tensor, so the free pressure behavior of a \emph{black hole universe} was foreseeable. By contrary,  this paper, based on  an approximate solution of the wave like equations (5), with an energy tensor with support on the  world lines of the particles, is not a \emph{black hole universe}. We consider it as a universe made of particles, representing macroscopic bodies without multipole structure.  
  \end{enumerate} 
  
  \vspace{0.5 cm}
  
 \appendix{\large\bf Appendix}

\begin{figure}[hbt]\label{A1}
\centering
\includegraphics[width=1.2\textwidth]{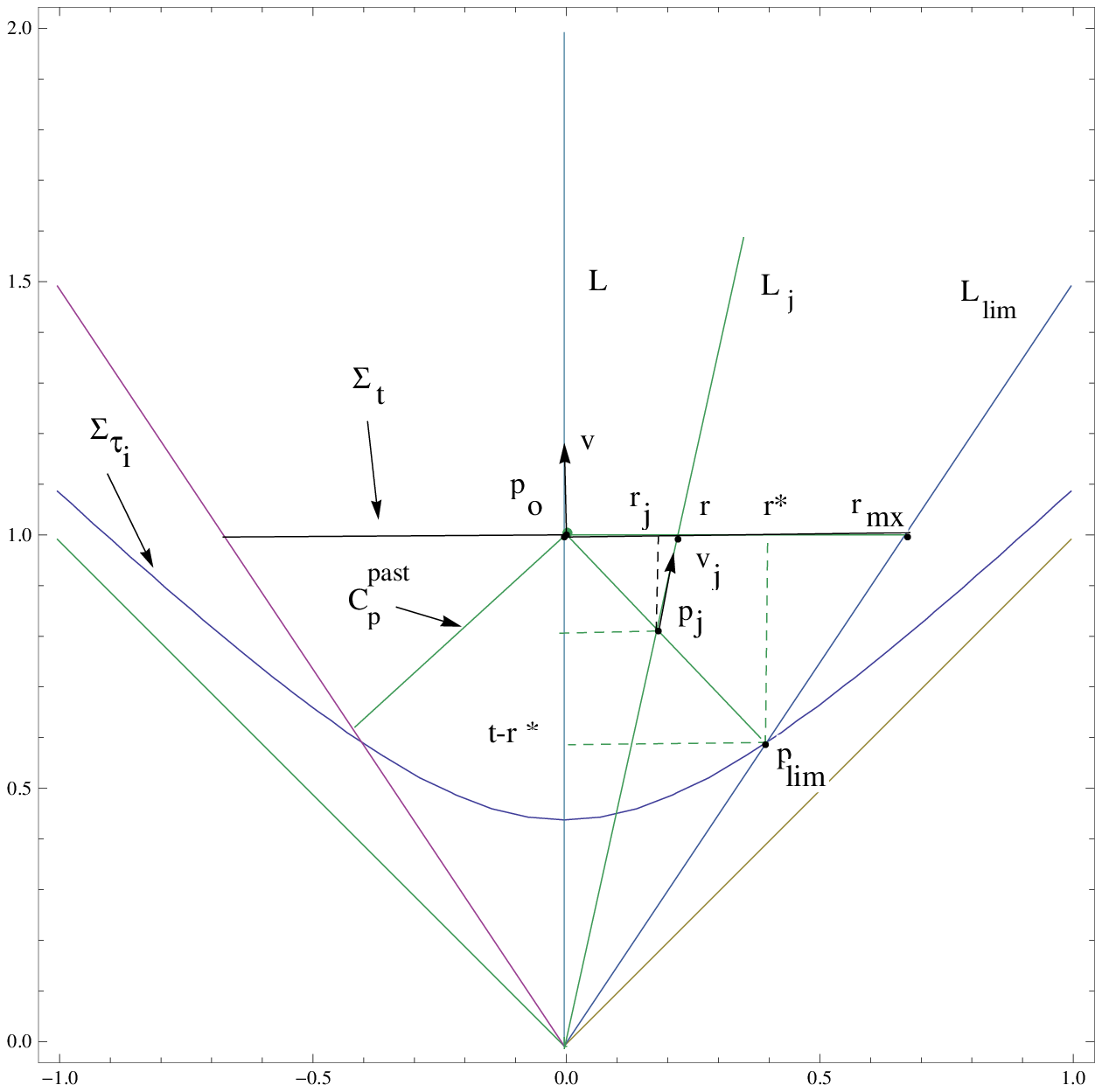}
\caption{This figure shows the meaning of the radial coordinates used in the appendix:}
\begin{enumerate}
\item $r_{mx}$ is the radial coordinate of the point $ L_{lim} \cap \Sigma_t$
\item $r^{*}$ is radial coordinate of the point $ L_{lim} \cap \Sigma_{\tau_i}$
\item $r_j$ is the radial coordinate of the point $ L_{j} \cap C^{past}_{p_o}$: the retarded position of particle $j$.
\item $r$ is the radial coordinate of the point $ L_{j} \cap \Sigma_t$
\end{enumerate}
\end{figure}

Let us show how we have got the statistical average of the   $g^{(1)}_{00}$ metric  component, given in equations (13-14). We must calculate the sum appearing in (9). We recall that the worldlines are parametrized with the Minkowskian time $\tau$. Let $p_o$ be a point on a world line of reference $L$, and choose coordinates $\{x^{\alpha}\}$ such that the tangent vector at $p_o$ to this line be $v= \partial_t$, and the  coordinates of  $p_o=(t,0,0,0)$. Let  $\Sigma_t= \{ x\in M / x^o=t=const.\}$, $\Sigma_{\tau}=  \{ x\in M / t^2-r^2= \tau=const.\}$, and denote by  $C^{past}_{p_o}$ the past light cone at $p_o$,  limited by the hiperboloid  $\Sigma_{\tau_i}$ which represents the birth of the galaxies at time $\tau_i$. The summands in (9) are referred to the retarded points $p_j$ where a particle world line $L_j$ intersects the  cone $C^{past}_{p_o}$. Let $r_j$ be its   radial coordinate and  $r$ the radial coordinate of the intersection point  $L_j \cap \Sigma_t $. Thus, the tangent vector to the world line $L_j$ at $p_j$ will be $v_j = \gamma_j (\partial_t +  \bar v_j\partial_r)$, with $\bar v_j =\frac{r}{t}$,  and $\gamma_j = t(t^2-r^2)^{-1/2}$. This is a unit vector with respect the Minkowski metric. With these notations we can write  $g^{(1)}_{00}(p_o)$  as follows:
\begin{equation}
g^{(1)}_{00}(p_o)= -G m \sum_{j} \frac{\gamma_j - 1/2}{r_j \gamma_j(1+\bar v_j)}
\end{equation}

 We have seen in section (3.1.1) that  a Milne Universe  has a uniform  number density over  hipersurfaces   $\Sigma_{\tau}$ of constant  Minkowski proper time, and in consequence   a non uniform density $ n^{(0)}(r,t)=Nt(t^2-r^2)^{-2}$,  on the hipersurfaces $\Sigma_t$. We shall make a statistical average by assuming a uniform random process over  $\Sigma_{\tau}$. This is equivalent to consider a point process on $\Sigma_t$ with the density  $ n^{(0)}(r,t)$ given above.
The statistical average is then given by the integral
\begin{equation}
\langle  g^{(1)}_{00}(p_o) \rangle= -4 G N  m \int_0^{r_{mx}} \frac{t}{(t^2-r^2)^2}\frac{\gamma_j^2 -1/2}{r_j \gamma_j (1+\bar v_j)} 4 \pi r^2 dr
\end{equation}
where $r_{mx}$ is the radial coordinate of the point intersection of $\Sigma_t$ with the world line $L_{lim}$ as shown in   Figure A1 . The set of worldliness like $L_{lim}$ defines the boundary of the particles that have interacted with the reference particle $L$ at time $t$. 

To perform the integral we need the relation between the radial coordinates $r_j$ and $r$.  It is straightforward to derive it from the Figure A1:  $r_j = \frac{r t}{r+t}$.
The result of the integration is
\begin{equation}\label{A1}
\langle  g^{(1)}_{00}(p_o) \rangle=-8\pi G N m \left(\frac{2}{3} t^2 (t^2-r^2_{mx})^{-3/2} - (t^2-r^2_{mx})^{-1/2}+\frac{1}{3t}\right)
\end{equation}
It remains to get the value of $r_{mx}$. We will shall show that it depends on both the present time $t$ and the initial time $\tau_i$. Let $r^{*}$ be the radial coordinate of the point $p_{lim}=  L_{lim} \cap \Sigma_t$. From the figure it is easy to get 
\begin{equation}
 r_{mx}= t \frac{r^{*}}{t-r^{*}}
 \end{equation}
 Next we solve the system $ \bar t +\bar r = t \ ,  \bar t^2 -\bar r^2 = \tau_i^2$ to get the radial coordinate $ r^{*}= \frac{t^2 -\tau_i^2}{2t}$, and substituting into the previous equation we get
 \begin{equation}
  r_{max}= t\frac{t^2-\tau_i^2}{t^2+\tau_i^2}
  \end{equation}
But, the intersection of the reference world line $L$ with the initial hipersurface $ \Sigma_{\tau_i}$ is a point with coordinate time $t_i=\tau_i $, thus we can write finally
 $r_{max}= t \frac{t^2-t_i^2}{t^2+t_i^2}$. It is straightforward to get the relations:
 \begin{equation}
 t^2 - r^2_{mx} =\frac{4 t^4 t_i^2}{(t^2+t_i^2)^2} \ , \  (t^2 - r^2_{mx})^{3/2} =\frac{8 t^6 t_i^3}{(t^2+t_i^2)^3} \ , \ ( t^2 - r^2_{mx})^{1/2} =\frac{2 t^2 t_i}{t^2+t_i^2}
 \end{equation}
 Substituting into the expression (\ref{A1}), and writing  $\mu_i = \frac{Nm}{t_i^3}$ for the initial mass density we get 
 \begin{equation}
 \langle  g^{(1)}_{00}(p_o) \rangle= -8\pi G \mu_i \left(\frac{1}{12} t^2 \left(1+\frac{t_i^2}{t^2}\right)^3- t_i^2\left(\frac{1}{2}\left(1+\frac{t_i^2}{t^2}\right)-\frac{1}{3}\frac{t_i}{t}\right)\right)
 \end{equation}

\noindent {\large\bf Acknowledgments}

\vspace{0.2cm}

\indent
 The author is indebted to his friends for their patience and to the support of the Spanish Ministry of \emph{Econom\'{\i}a y Competitividad}, MICINN-FEDER project FIS2012-33582.

\end{document}